\documentclass[12pt,preprint]{aastex}
\usepackage{amssymb}
\usepackage{graphicx}

\shorttitle{The Xinglong  2.16-m Telescope}
\shortauthors{Fan et al.}

\begin{document}

\title{The Xinglong 2.16-m Telescope: Current Instruments and Scientific Projects}

\author{Zhou Fan\altaffilmark{1}, Huijuan Wang\altaffilmark{1}, 
  Xiaojun Jiang\altaffilmark{1}, Hong Wu\altaffilmark{1},
  Hongbin Li\altaffilmark{1}, Yang Huang\altaffilmark{2},  
  Dawei Xu\altaffilmark{3}, Zhongwen Hu\altaffilmark{4},
  Yinan Zhu\altaffilmark{1}, Jianfeng Wang\altaffilmark{1},
  Stefanie Komossa\altaffilmark{5}, Xiaoming Zhang\altaffilmark{1}}

\altaffiltext{1}{Key Laboratory of Optical Astronomy, National Astronomical
  Observatories, Chinese Academy of Sciences, 20A Datun Road, Chaoyang
  District, Beijing 100012, China}

\altaffiltext{2}{Department of Astronomy, Peking University, Beijing
  100871, China}

\altaffiltext{3}{Key Laboratory of Space Astronomy and Technology, National 
  Astronomical Observatories, Chinese Academy of Sciences, 20A Datun Road, 
  Chaoyang District, Beijing 100012, China}

\altaffiltext{4}{Nanjing Institute of Astronomical Optics \& Technology, 
  National Astronomical Observatories, Chinese Academy of Sciences,
  188 Bancang Street, Nanjing 210042, China}

\altaffiltext{5}{Max-Planck-Institut fuer Radioastronomie, Auf dem Huegel 69,
  53121 Bonn, Germany}

\email{zfan@bao.ac.cn}

\begin{abstract}
The Xinglong 2.16-m reflector is the first 2-meter class astronomical 
telescope in China. It was jointly designed and built by the Nanjing 
Astronomical Instruments Factory (NAIF), Beijing Astronomical 
Observatory (now National Astronomical Observatories, Chinese Academy of 
Sciences, NAOC) and Institute of Automation, Chinese Academy of 
Sciences in 1989. It is Ritchey-Chr\'{e}tien (R-C) reflector on an English 
equatorial mount and the effective aperture is 2.16 meters. It had been 
the largest optical telescope in China for $\sim18$ years until the
Guoshoujing Telescope (also called Large Sky Area Multi-Object Fiber 
Spectroscopic Telescope, LAMOST) and the Lijiang 2.4-m telescope were 
built. At present, there are three main instruments on the Cassegrain 
focus available: the Beijing Faint Object Spectrograph and Camera (BFOSC) for 
direct imaging and low resolution ($R\sim500-2000$) spectroscopy, the 
spectrograph made by Optomechanics Research Inc. (OMR) for low resolution 
spectroscopy (the spectral resolutions are similar to those 
of BFOSC) and the fiber-fed High Resolution Spectrograph 
(HRS, $R\sim30000-65000$). The telescope is widely open to astronomers 
all over China as well as international astronomical observers. 
Each year there are more than 40 ongoing observing projects, 
including 6-8 key projects. 
Recently, some new techniques and instruments (e.g., astro-frequency comb 
calibration system, polarimeter and adaptive optics) have been 
or will be tested on the telescope to extend its observing abilities.
\end{abstract}

\section{Introduction}
\label{intro.sec}

The 2.16-m reflector is an English equatorial mount telescope at Xinglong 
Observatory and it obtained the first light in 1989, which was 
built/developed by Chinese ourselves independently. 
The effective aperture of the telescope is 2.16 meter, and the focal ratio of 
the primary mirror is f/3 \citep{szy89}. The effective aperture of the 
secondary mirror is 0.717 m. Currently, there are two available foci for 
mounting astronomical observing instruments, the Cassegrain focus and the 
Coud\'{e} focus. For the Cassegrain focus, which is a R-C system, the focal 
ratio is f/9, and the scale on the focal plane is $10\farcs61$/mm. While for 
the Coud\'{e} system, it is f/45, and the scale on the focal plane is 
$2\farcs12$/mm. 
The most special character of the optical system is that the Cassegrain 
system and the Coud\'{e} system share the same secondary mirror and there is 
a relay mirror in the Coud\'{e} system, and both systems can 
sufficiently eliminate the spherical aberration and coma aberration 
\citep{szy89}.

Since its first light in 1989, the 2.16-m telescope has been used in various 
scientific research fields in Galactic and extragalactic astrophysics, 
including the determination of stellar parameters (e.g., 
abundances, surface gravity, temperatures) of a large sample of stars, 
the discoveries of substellar and planetary companions of stars, studies of 
active galaxy nuclei (AGN), including the identifications of high-redshift 
quasars, discoveries and studies of supernovae, e.g., 1993J \citep{wh94},
as well as time-domain science (e.g., supernovae, gamma-ray bursts, 
stellar tidal disruption events, and variable stars). 

Although the  2.16-m telescope is the third largest optical telescope in
China right now (smaller than LAMOST and the Lijiang 2.4-m telescopes, which 
were installed in 2008 and 2007, respectively), it plays an important role in 
Chinese astronomical observations. Every year there are hundreds of Chinese 
astronomers applying for observing time at the 2.16-m telescope, including 
significant numbers of new telescope users and graduate students. 
Therefore, it is of 
great importance and useful to describe the specific parameters and the
observing ability of the telescope and its instruments to the telescope users,
which could be very helpful for planning and carrying out observations
and for the data reduction. This paper is 
organized as follows: the Xinglong Observatory is introduced in 
Sect.\ref{obs.sec}; the telescope and its instruments are described in 
Sect.\ref{tel.sec} and the efficiencies of the telescope and its instruments 
also have been estimated in this section. The relevant science and
projects based on observational data obtained by the 2.16-m telescope are 
described in Sect.\ref{sci.sec}; finally a brief summary is given and some
future facilities are discussed in Sect.\ref{sum.sec}.

\section{The Xinglong Observatory}
\label{obs.sec}

The Xinglong Observatory of National Astronomical Observatories, CAS (NAOC) 
(IAU code 327, coordinates: $40^{\circ}23'39''$ N, $117^{\circ}34'30''$ E) 
was founded in 1968. At present, it is one of most primary observing stations 
of NAOC. As the largest optical astronomical observatory site in the continent 
of Asia, it harbors 9 telescopes with effective aperture greater 
than 50\,cm. These are LAMOST, the 2.16-m reflector, a 1.26-m optical \& 
near-infrared reflector, a 1\,m Alt-Az reflector, a 85\,cm reflector
(NAOC-Beijing Normal University Telescope, NBT), a 80\,cm reflector (Tsinghua
University-NAOC, 
TNT), a 60/90\,cm Schmidt telescope, a 60\,cm reflector and a 50\,cm reflector. 
The average altitude of the Xinglong Observatory is $\sim960$ m and it is
located at the South of the main peak of the Yanshan Mountains, in the 
Xinglong county, Hebei province, which is $\sim120$km northeast of Beijing. 
The mean and median seeing values of the Xinglong Observatory are 
$2\farcs0$ and $1\farcs8$, respectively, and on average, there are 117 
photometric nights and 230 useful nights per year based on the data of 
2007-2014 \citep{zhang15}. For most of the time, the wind speed is less 
than 4 m s$^{-1}$ (the mean value is 2 m s$^{-1}$), and the sky 
brightness is $\sim$21.1 mag arcsec$^{2}$ in $V$ band at the zenith 
\citep{zhang15}.

Each year, more than a hundred astronomers use the telescopes of 
Xinglong Observatory to perform the observations for the studies on Galactic 
sciences (stellar parameters, extinction measurements, Galactic 
structures, exoplanets, etc.) and extragalactic sciences (including nearby
galaxies, AGNs, high-redshift quasars), as well as time-domain astronomy 
(supernovae, gamma-ray bursts, stellar tidal disruption events, and different
types of variable stars). 
In recent year, besides the basic daily maintenance of the telescopes, 
new techniques and methods have been explored by the engineers and 
technicians of Xinglong Observatory to improve the efficiency of observations.
Meanwhile, the Xinglong Observatory is also a National popular-science and 
education base of China for training students from graduate schools, 
colleges, high schools and other education institutions throughout China,
and it has hosted a number of international workshops and summer schools. 

\section{The  2.16-m reflector and its instruments}
\label{tel.sec}

\begin{figure}
  \centerline{
    \includegraphics[scale=0.5,angle=-90]{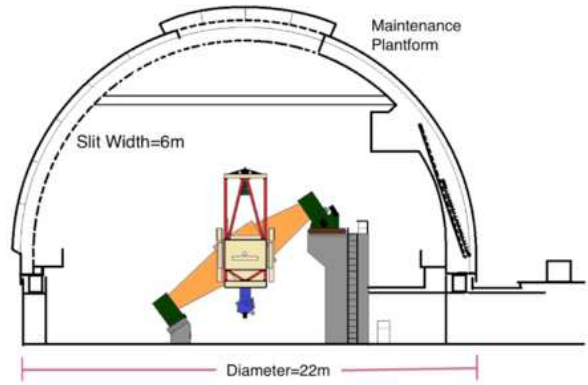}}
  \caption[]{Cartoon of the  2.16-m telescope at Xinglong Observatory, 
    which is an English equatorial mount, in a large dome with diameter of 22 
    meter.}
  \label{fig1}
\end{figure}

\begin{figure}
  \centerline{
    \includegraphics[scale=0.4,angle=0]{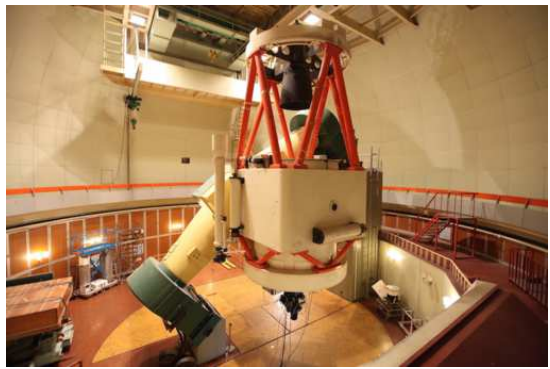}}
  \caption[]{The Xinglong 2.16-m telescope at Xinglong Observatory.}
  \label{fig2}
\end{figure}

As shown in Figures~\ref{fig1} and \ref{fig2}, the telescope is a R-C system
on an English equatorial mount. The effective aperture of the primary mirror
is 2.16 meter. Currently, there are three primary instruments
available for the Cassegrain focus: BFOSC, OMR and the fiber-fed HRS.
Previously,  
a high resolution spectrograph was mounted on the Coud\'{e} focus before the 
fiber-fed HRS was mounted in 2010. Since then, the Coud\'{e} focus has   
only been used for a few special experiments, such as adaptive optics 
tests. As shown in Figure~\ref{fig1}, the diameter of the dome is 22 meter,  
which is relative large for a 2m-class telescope, due to its design. The 
rotation part of the telescope weights 91 tons and the pointing accuracy 
which is modified with pointing model is $rms=10\farcs0$. While the 
tracking precision of the telescope is $rms=20\farcs0$ within 10 minutes, and 
$rms=50\farcs0$ within 1 hour. The position precision of the telescope on the 
sky under guiding is $rms<0\farcs15$ \citep{huang15}.

\subsection{The BFOSC instrument}

\begin{figure}
  \centerline{
    \includegraphics[scale=0.8,angle=0]{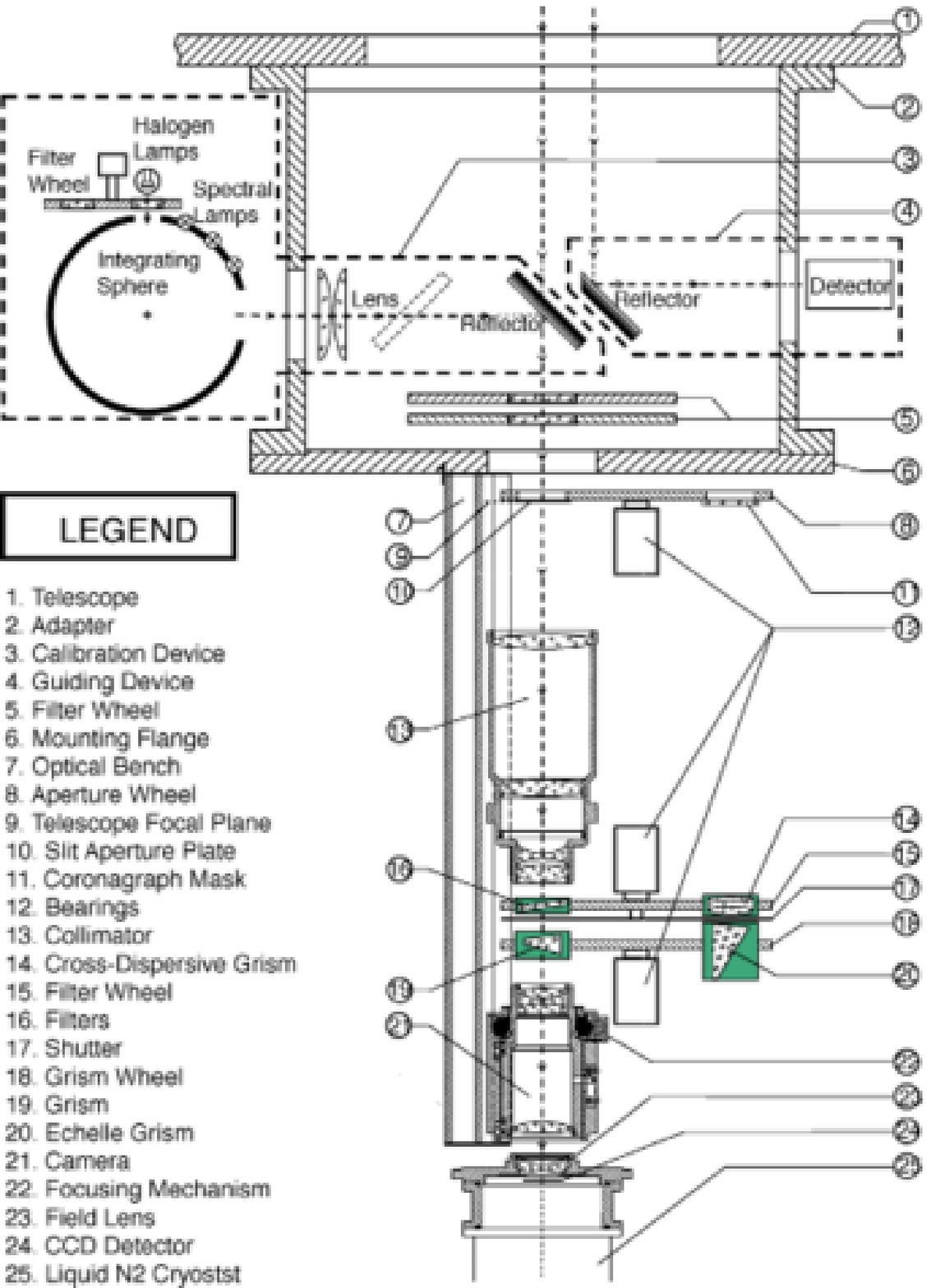}}
  \caption{The optical and mechanical layout of the BFOSC instrument 
    from \citet{hwl}.}
  \label{fig3}
\end{figure}

The BFOSC is one of the primary instruments of the telescope, for which
the design and processing of mechanism and electronics control, the design of
grims and assembling and debugging of the whole system were done by
University of Copenhagen and ESO provided consult and optical design of the
focal reducer. It is available for the f/9 Cassegrain focus. The scale of
the focal plane is 
$10\farcs61$/mm. As discussed in Sect.~\ref{intro.sec}, it can 
be used for both imaging mode and spectroscopy mode, which is switchable. 
Figure~\ref{fig3} shows the optical layout of the BFOSC instrument, including 
the aperture wheel, filter wheel, grism/echelle wheel, calibration device, 
guiding device, collimator, shutter, camera and the CCD detector.

There are eight positions on the aperture wheel: three of them for direct 
imaging, coronagraph mask and focal adjusting plate, respectively and the 
other five for long/short slit plates. For the long slits, the lengths are
all $9'.4$ and there are nine options for the slit widths ($0\farcs6$, 
$0\farcs7$, $1\farcs1$, $1\farcs4$, $1\farcs8$, $2\farcs3$, $3\farcs6$,
$7\farcs0$ and $14\farcs0$). While for the 
short slits, the slit lengths are various for different slit width: 
$3\farcs5$ for slit width of $0\farcs6$, $4\farcs0$ for slit width of 
$1\farcs0$, $3\farcs6$ for slit width of $1\farcs6$ and $3\farcs7$ for 
both slit widths of $2\farcs3$ and $3\farcs2$. The coronagraph mask is used 
to block out the intense light from bright sources near the observing targets,
with circular spots of which the diameters are $2\farcs0$, $3\farcs0$, 
$4\farcs0$, $6\farcs0$, $9\farcs0$ and $12\farcs0$.  

\begin{figure}
  \centerline{
    \includegraphics[scale=0.7,angle=-90]{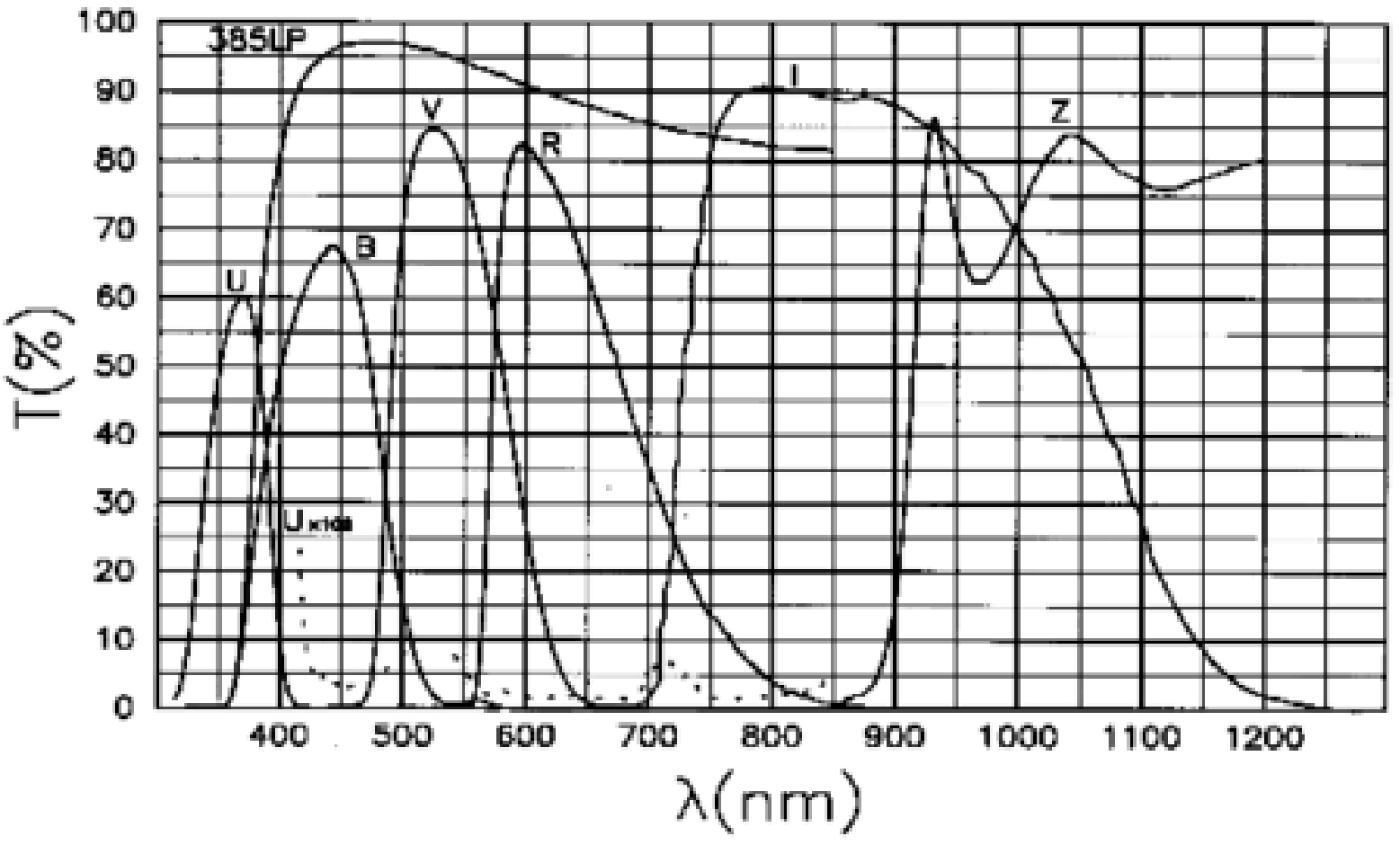}}
  \caption[]{The transmission curves of the BFOSC filters, including the
    Johnson-cousins $UBVRI$ bands as well as the $385LP$ and $Z$ bands for
    spectroscopic observations.}
  \label{fig4}
\end{figure}

For the filter wheel, there are several sets of filters available. The 
Johnson-Cousins $UBVRI$ filters are for the broadband photometry. 
In addition, there are special filters used for the spectroscopic observing: 
the $Z$ band for the transmitting spectral region $\lambda\ge910$ nm, and the 
$385LP$ band for removing the 2nd-order spectrum in the wavelength region 
$\lambda\le385$ nm. The transmission curves are shown in Figure~\ref{fig4}.

Besides, a series of interference narrow band filters covering [OIII] 
(rest-frame wavelength of $\lambda$5007A), $\rm H{\alpha}$ and HeII 
(rest-frame wavelength $\lambda$4686A) are also available for the BFSOC 
imaging mode. For series of [OIII] band observations, there are 8 filters of
which the central wavelengths are
between 500.9-536.0 nm with FWHM of 6 nm, corresponding to redshifts of
0-20000 km s$^{-1}$. While for the series of $\rm H_{\alpha}$ bands, there
are 11 filters of which the central wavelengths are between 656.2-706.0 nm
with FWHM of 7 nm, corresponding to redshifts of 0-22000 km s$^{-1}$. For
the He bands, there are three filters of which the central wavelengths are
between 447.1-468.6 nm with FWHM of 6 nm. In addition, an [OIII] filter with
FWHM of 13 nm and an $\rm H_{\alpha}$ filter with FWHM of 14 nm are also
provided for a redshift of z=0. These filters can be used for the study
of star-forming regions of nearby galaxies at different redshifts.

Table~\ref{t1.tab} presents the parameters of the grisms/prisms/echelle for
the BFOSC instrument. From left to right in columns are names, working spectral
orders, reciprocal linear dispersions, dispersions and wavelength coverages of
the various grisms/prisms/echelle, respectively. The configurations can be
chosen by the users depending on the different requirements of the projects.
For grisms G3/G6/G10, the blue ends of wavelength coverages are limited by the 
cutoff of the atmospheric window, while for grisms G5/G8, the red ends are
limited by the size of the CCD. The low dispersion grisms G10/G11/G12 are also
used for the cross-disperser when mounted on the filter wheel. The echelle E13
is only used for measuring the velocity field of extended sources with the 3rd
order spectrum, and the $V$ band filter is recommended to be applied together
with to remove the other order of the spectrum.

\begin{table}[ht!!!]
\small
\centering
\begin{minipage}[]{100mm}
  \caption[]{The Parameters of Grisms/Prisms/Echelle for the BFOSC Instrument
    from \citet{hwl}.}
\label{t1.tab}
\end{minipage}
\tabcolsep 3mm
\begin{tabular}{ccccc}
  \hline\noalign{\smallskip}
  Name & Spec.Ord. & Rec. Lin. Disp. & Disp. & Wav. Range \\
  &  & (\AA/mm) & (\AA/pix) & (\AA) \\
  \hline\noalign{\smallskip}
  P1  &   & $573-2547$ & $8.6-38.2$ & $4000-5600$ \\
  \hline
  G3  & 1 & 139 & 3.12  & $3300-6400$ \\
  G4  & 1 & 198 & 4.45  & $3850-7000$ \\
  G5  & 1 & 199 & 4.47  & $5200-10120$ \\
  \hline
  G6  & 1 & 88  & 1.98 & $3300-5450$ \\
  G7  & 1 & 95  & 2.13 & $3870-6760$ \\
  G8  & 1 & 80  & 1.79 & $5800-8280$ \\
  \hline
  G10  & 1 & 392  & 8.80 & $3300-6400$ \\
  G11  & 1 & 295  & 6.63 & $3900-7300$ \\
  G12  & 1 & 837  & 18.8 & $5200-10200$ \\
  \hline
  E9+G10  & $21-11$ & $16.8-38.4$  & $0.38-0.86$ & $3300-6400$ \\
  E9+G11  & $18-9$ & $21.0-47.9$  & $0.47-1.076$ & $3900-7300$ \\
  E9+G12  & $14-6$ & $29.0-73.2$  & $0.65-1.64$  & $5200-10200$ \\
  \hline
  E13+V  & 3 & 33.1  & 0.666  & $4980-5990$ \\
  \noalign{\smallskip}\hline
  \noalign{\smallskip}\hline
\end{tabular}
\end{table}

Table~\ref{t2.tab} presents the spectral resolutions of some frequently used
BFSOC grisms at a minimum slit width $0\farcs6$ and a slit width of 
  $\sim2\farcs3$ used in typical seeing condition, which are estimated with
the emission lines of the planetary nebula (PN) IC4997. The observations were
taken on November 24, 2014 and February 26, 2016, with gratings of G4, G6, 
G7 and G8 on BFOSC. Throughout the nights of the observations, the weather were
clear and the seeing were between $\sim2\farcs0 - 2\farcs4$. The exposure time 
was 0.2-10 seconds depending on the dispersions of the different grisms. 
The shortest exposure of 0.2 second is considerably longer than the shutter
speed of 1.5 milliseconds.

\begin{table}[ht!!!]
\small
\centering
\begin{minipage}[]{100mm}
  \caption[]{Spectral resolutions of several frequently used grisms of BFSOC at\
 a slit width of $0\farcs6$ and $2\farcs3$, which is around the typical seeing.}
  \label{t2.tab}
\end{minipage}
\tabcolsep 3mm
\begin{tabular}{c|cccc|cccc}
  \hline\noalign{\smallskip}
  
  
  
  Wavelength & G4 & G6 & G7 & G8 & G4 & G6 & G7 & G8 \\
  \hline\noalign{\smallskip}
   ({\AA})    & \multicolumn{4}{c|}{$0\farcs6$}     & \multicolumn{4}{c}{$2\farcs3$} \\
  \hline\noalign{\smallskip}
  4341  & 531  & 1423 & 1204 &  \nodata & 246 &463 &478 & \nodata \\
  4363 & 557 & 1503 & 1252 & \nodata  & \nodata &\nodata &\nodata & \nodata \\ 
  4861  & \nodata &\nodata &\nodata &\nodata &    262   & 538    &   521  & \nodata \\
  4959  &       &          &         &          & 265  & 540 & 524 &  \nodata \\
  5007 & 620 & 1413 & 1328 &  \nodata & 265  & 555 & 532  & \nodata \\
  6563 & 824 & \nodata &   \nodata &  2245  & 321 &  \nodata  &  \nodata  & 820 \\
  
  \noalign{\smallskip}\hline
\end{tabular}
\end{table}

For the calibration device, four lamps can be mounted at most. One is for 
the flat-fielding correction and the other lamps can be used for wavelength 
calibration. The lights of the lamps illuminate the integrating sphere at 
first, then modified to be f/9 light beam and finally reflected to the 
focal plane of the BFOSC instrument. The Fe lamp and Ne lamp are frequently 
used for the wavelength calibration.  

On the guiding device, a mirror reflects the view outside of the observing 
field of view (FOV) to an ICCD, which is movable in 3D directions to find a 
proper guide star and also to adjust the focus. The FOV of the guiding device 
is $10'\times20'$, which is large enough to find a suitable guide star.

\begin{figure}
  \centerline{
    \includegraphics[scale=0.7,angle=0]{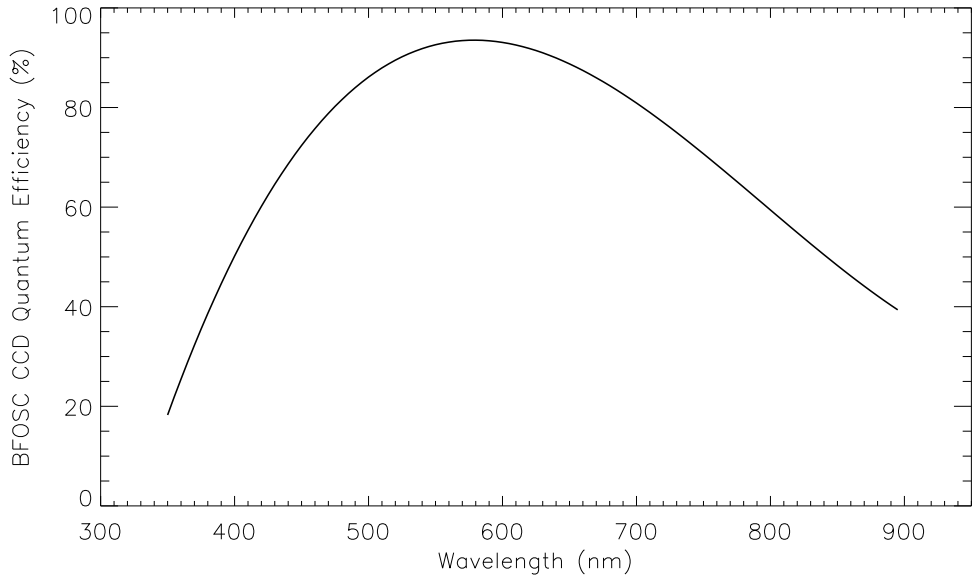}}
  \caption[]{The quantum efficiency of the BFOSC CCD camera.}
  \label{fig5}
\end{figure}

In 2010 an E2V $55-30-1-348$ back illuminated CCD, AIMO was installed on the
spectrograph and the CCD controller was made by the Lick Observatory. The
size of the CCD is $1242\times1152$ pixels with pixel size of 22.5 $\mu$m.
The pixel scale is $0\farcs457$ and the FOV is $9'.46 \times8'.77$ according 
to the size of CCD. Figure~\ref{fig5} shows the quantum efficiency (Q.E.) of 
the CCD. It can be seen that the maximum quantum efficiency is higher than 
90\% around 5700 {\AA} of the wavelength.

Table~\ref{t3.tab} lists the gains, readout noises and readout speed of
the BFOSC CCD at various readout times. Observers can choose different options
for the specific observations and the slow readout speed is applied in
most of the time.

\begin{table}[ht!!!]
\small
\centering
\begin{minipage}[]{100mm}
  \caption[]{Gains, Readout Noises and readout Speed of BFOSC CCD from
    \citet{hwl}.}
  \label{t3.tab}\end{minipage}
\tabcolsep 3mm
\begin{tabular}{ccccc}
  \hline\noalign{\smallskip}
  Readout Speed & Mode & Gain & Readout Noise & Readout Time \\
  &   & ($e^-$/ADU) & ($e^-$/pix.) & (second) \\
  \hline\noalign{\smallskip}
  Fast   &  0  & 99.58 & 132.5 & 5  \\
  Fast   &  1  & 49.44 & 103.7 & 5  \\
  Fast   &  2  & 24.48 & 87.35 & 5  \\
  Median &  0  & 2.33 & 6.80 & 8  \\
  Median &  1  & 1.22 & 5.93 & 8  \\
  Median &  2  & 0.50 & 6.06 & 8  \\
  Slow   &  0  & 2.43 & 3.25 & 28  \\
  Slow   &  1  & 1.13 & 2.58 & 28  \\
  Slow   &  2  & 0.50 & 2.46 & 28  \\
  \noalign{\smallskip}\hline
\end{tabular}
\end{table}

\begin{figure}
  \centerline{
    \includegraphics[scale=0.9,angle=0]{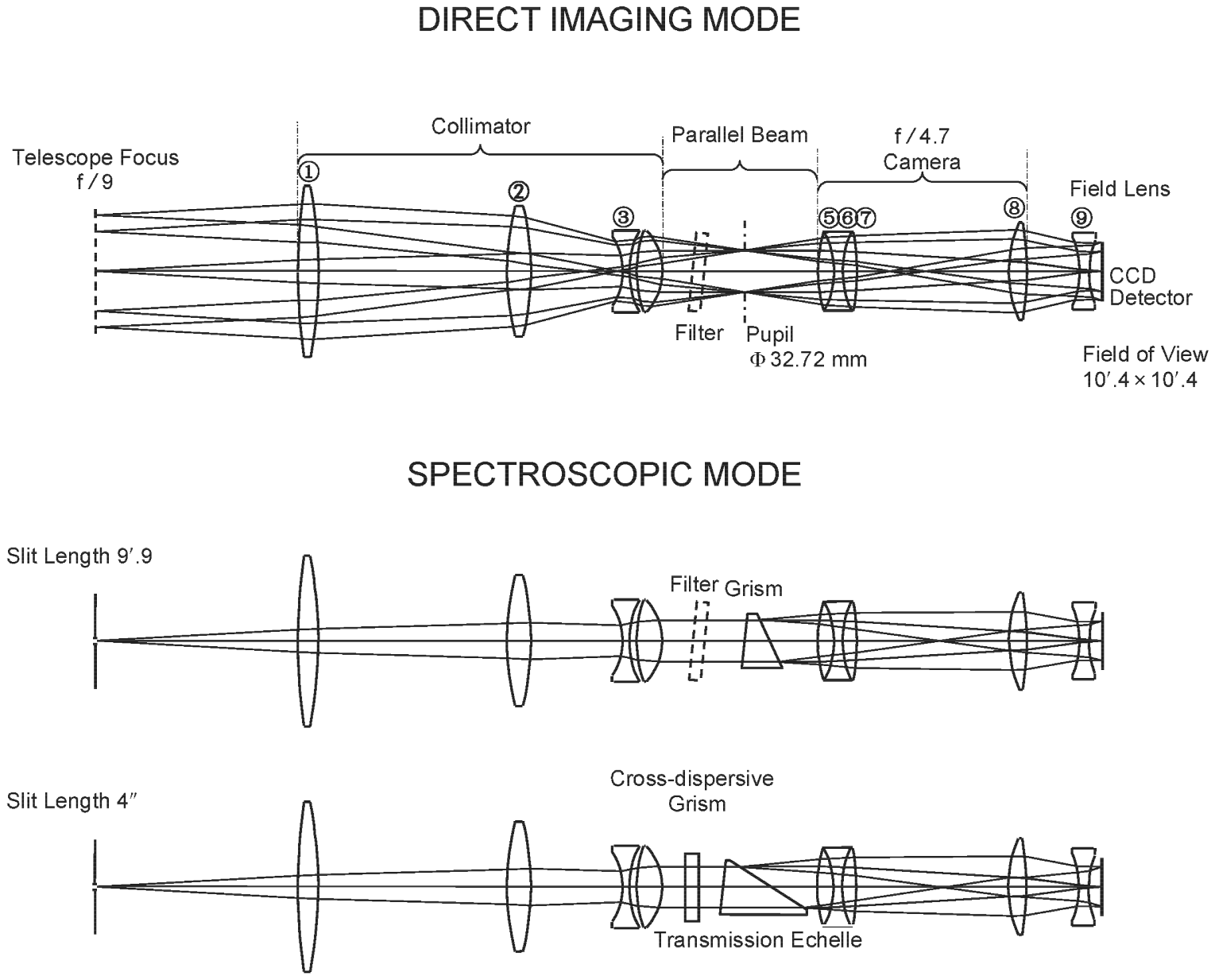}}
  \caption{The layout of the lens of two different working modes of the BFOSC
    instrument from \citet{hwl}.}
  \label{fig6}
\end{figure}

There are six observing modes for the BFOSC instrument:
(1) direct imaging; (2) long slit spectroscopy; (3) slitless
spectroscopy; (4) echelle grism spectroscopy; (5) coronograph mask;
and (6) multiple object Spectroscopy. Figure~\ref{fig6} shows the
optical layout of the lens and light path for the direct imaging mode
and the spectroscopy observing mode, which could be switched in a few minutes.

Since 2012, a multi-object spectroscopy (MOS) observing mode of BFSOC is 
available for observers by placing a multiple-aperture mask on the 
aperture wheel \citep{zhou14}. In this mode, 10-20 objects can be observed 
simultaneously, depending on the spatial distribution density of the targets. 
The MOS improves the observing efficiency of multiple-object observations, 
such as star-forming activities of HII regions of nearby galaxies, 
star clusters, or groups and clusters of galaxies.

The total efficiency of the observing system, including the atmosphere, 
telescope and its instruments, is what the observers are mostly concerned 
about. In fact, there are a number of factors to be considered for calculating 
the total efficiency of the system, such as 
atmospheric extinction, reflectivity of primary and secondary mirror, 
transmissions of filters and the quantum efficiency of the CCD. It can be 
estimated through observing standard stars in the Equation~\ref{eq1},

\begin{equation}
\label{eq1}
\eta (\lambda)=\frac{F_{ADU} \cdot G}{F_{\lambda} \cdot \delta \lambda \cdot S_{tel}}
\end{equation}

In this formula, $F_{\rm ADU}$ is the observed number counts of a standard 
star per second (ADU s$^{-1}$); $G$ is the gain of the CCD ($e^{-1}$ ADU$^{-1}$); 
$F_{\lambda}$ is theoretical photon flux of a standard star derived from its 
AB mag ($photon s^{-1} cm^{-2}$ \AA$^{-1}$) ; $\delta \lambda$ is the 
effective band width of the filter in imaging observations or the dispersion of 
the grating for spectroscopic observations (\AA); 
$S_{\rm tel}$ is the effective area of the primary mirror of the telescope 
($cm^2$) and $\lambda$ is the effective wavelength of the filter or the 
wavelength at which the efficiency is to be computed for the spectroscopy (\AA).

\begin{figure}
  \centerline{
    \includegraphics[scale=0.5,angle=0]{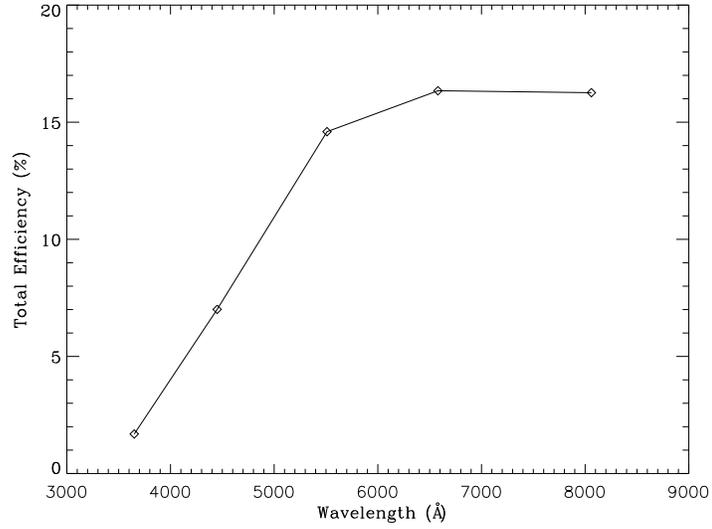}}
  \caption[]{The total efficiency of the BFOSC photometric system including the atmosphere of the 2.16-m telescope in the $UBVRI$ bands at an airmass of $\sim1.3$.}
  \label{fig7}
\end{figure}

In order to estimate the total efficiency of the  2.16-m telescope, a Landolt 
standard star PG2336+004B was observed on November 21, 2014 in the broadband 
$UBVRI$ bands of the BFOSC photometric system by using Equation~\ref{eq1}. 
The seeing was $\sim2\farcs0-3\farcs0$ throughout the whole night of 
observations and the airmass is $\sim1.3$. Figure~\ref{fig7} is a plot of
the total efficiency in different 
bands as a function of central wavelength. The atmospheric extinction, 
reflectivity of primary and secondary mirrors, transmissions of filters,
Q.E. of CCD as well as other factors were not corrected in the calculation. 
It can be seen that the total efficiency is relative low in the $U$ band 
($\sim2$\%) and $B$ band ($\sim7$\%), but it is relatively high in $VRI$ bands
($\sim15$\%).

\begin{figure}
  \centerline{
    \includegraphics[scale=0.5,angle=0]{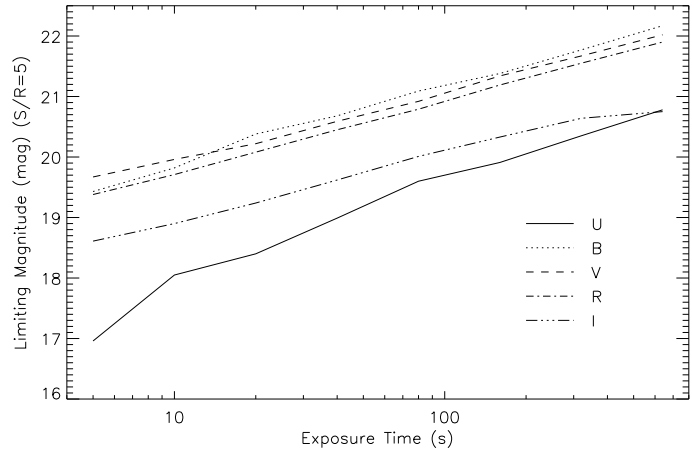}}
  \caption[]{The limiting magnitudes in the $UBVRI$ bands for the 
    signal-to-noise ratio of $SNR=5$ at an airmass of $\sim1.3$.}
  \label{fig8}
\end{figure}

\begin{figure}
  \centerline{
    \includegraphics[scale=0.5,angle=0]{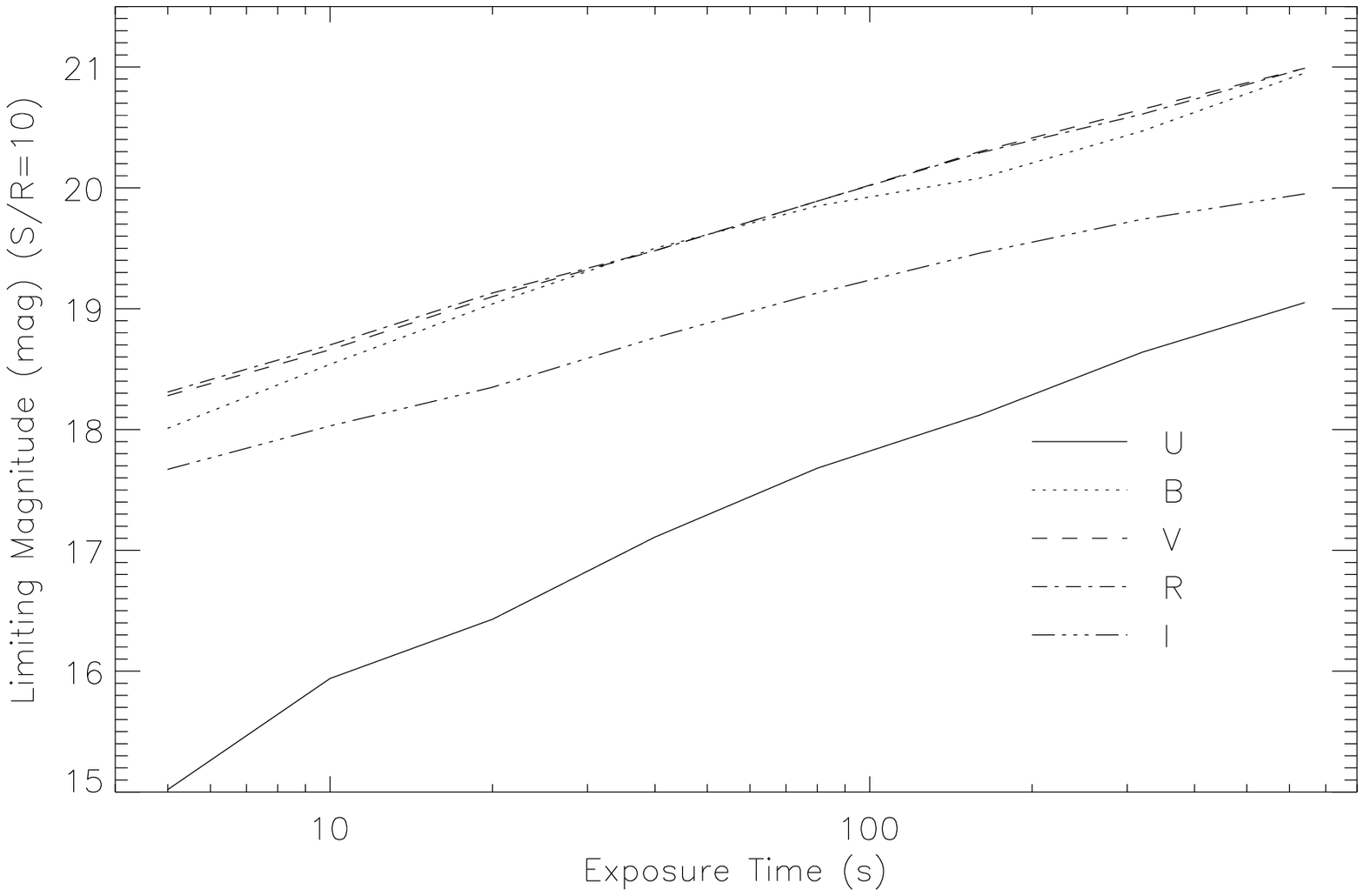}}
  \caption[]{The same as in Figure~\ref{fig8}, but for a signal-to-noise ratio 
    of $SNR=10$.}
  \label{fig9}
\end{figure}

The limiting magnitude is also an important quantity to evaluate the observing 
ability of a telescope. Actually at the same night on November 21, 2014,
a Landolt standard star PG2336+004 was observed in the $UBVRI$ bands. The 
airmass was  $\sim1.3$. The limiting magnitude (signal-to-noise 
ratio, $SNR=5$) as a function of exposure time in the $UBVRI$ bands are shown 
in Figure~\ref{fig8} and the same relations but for the limiting magnitudes 
of $SNR=10$ are shown in Figure~\ref{fig9}. It is noted that in the 640s 
exposure the limiting magnitudes of the $B/V/R$ bands could reach $\sim22$ mag 
in $SNR=5$ and $\sim21$ mag in $SNR=10$. The limiting magnitude in the $U$ 
band is the lowest because the sensitivity of the observing system in the blue 
band is relatively low, further affected by the effect of significant 
atmospheric extinction in blue band. However, although the efficiency in the 
$I$ band is similar to those of the $VR$ bands, the night-sky background is 
much brighter ($\sim2$ times) than in the $VR$ bands. In addition, the fringe 
of the CCD in the $I$ band contributes significant noise. Although we have 
tried the de-fringing method introduced by \citet{c87}, it has not been 
improved significantly. Therefore, these factors together make the limiting 
magnitudes $\sim1$ mag shallower than those in the $BVR$ bands.

\begin{figure}
\centerline{
  \includegraphics[scale=0.55,angle=0]{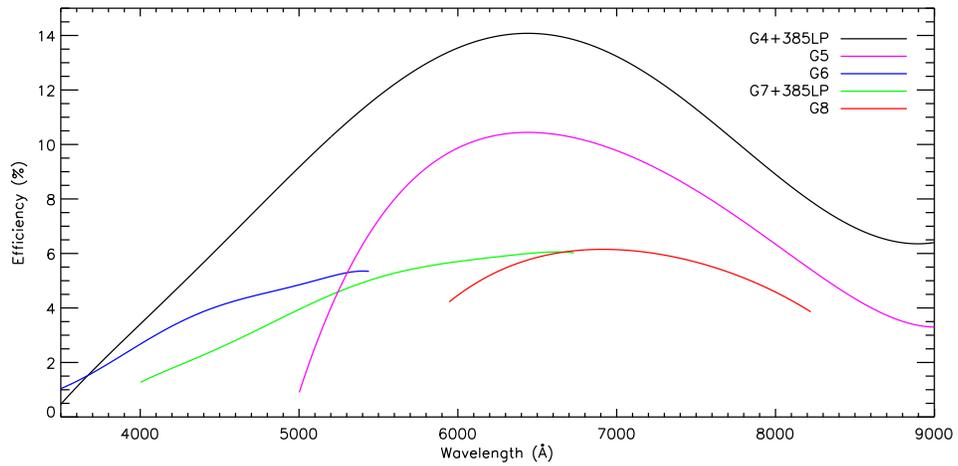}}
\caption[]{The total efficiencies estimated for grisms G4, G5, G6, G7 and G8 
  of the BFOSC instrument.}
\label{fig10}
\end{figure}

In order to estimate the efficiency of the spectroscopic system of BFOSC, 
the two ESO standard stars HR153 and HR9087 were observed on November 24, 
2014, with several frequently used grisms (G4, G5, G6, G7 and G8)
of the BFOSC instrument. The weather was clear and the seeing was 
$\sim2\farcs4$. The exposure time was 0.2-10 seconds according to the 
dispersions of the grisms. 
To make sure that most of the flux can be obtained and measured for 
estimating the efficiency, the slit was configured as $7\farcs0$. 
The total efficiency, which is defined above, including 
atmospheric extinction and instruments, was calculated with
Equation~\ref{eq1} and shown in Figure~\ref{fig10}. For the grisms G4 and G5,
the peak efficiencies are $\sim$15\% and $\sim$10\% respectively,
while for the other three grisms, the total efficiencies of the peak
are $\sim$5\%. The filter $385LP$ was used for removing the 2nd-order 
spectrum of wavelength $\lambda \ge$385 nm.

For the limiting magnitude in the spectroscopic observations, the 
previous observations show that typically it could reach $V=20$ mag of 
$SNR=5$ in the grism G6, with exposure of 1 hour, given by \citet{hwl}.

\subsection{The Spectrograph made by Optomechanics Research inc. (OMR)}

The OMR is another low-intermediate resolution spectrograph available on
the Cassegrain focus of the  2.16-m telescope. It was made by Optomechanics 
Research Inc. (Arizona, USA) at the end of 1994 and was tested on the 2.1m 
telescope at Kitt Peak. Then, it was installed on the  2.16-m telescope of 
Xinglong Observatory in 1995 and was available to astronomers in 1996. 

\begin{figure}
\centerline{
  \includegraphics[scale=0.9,angle=-90]{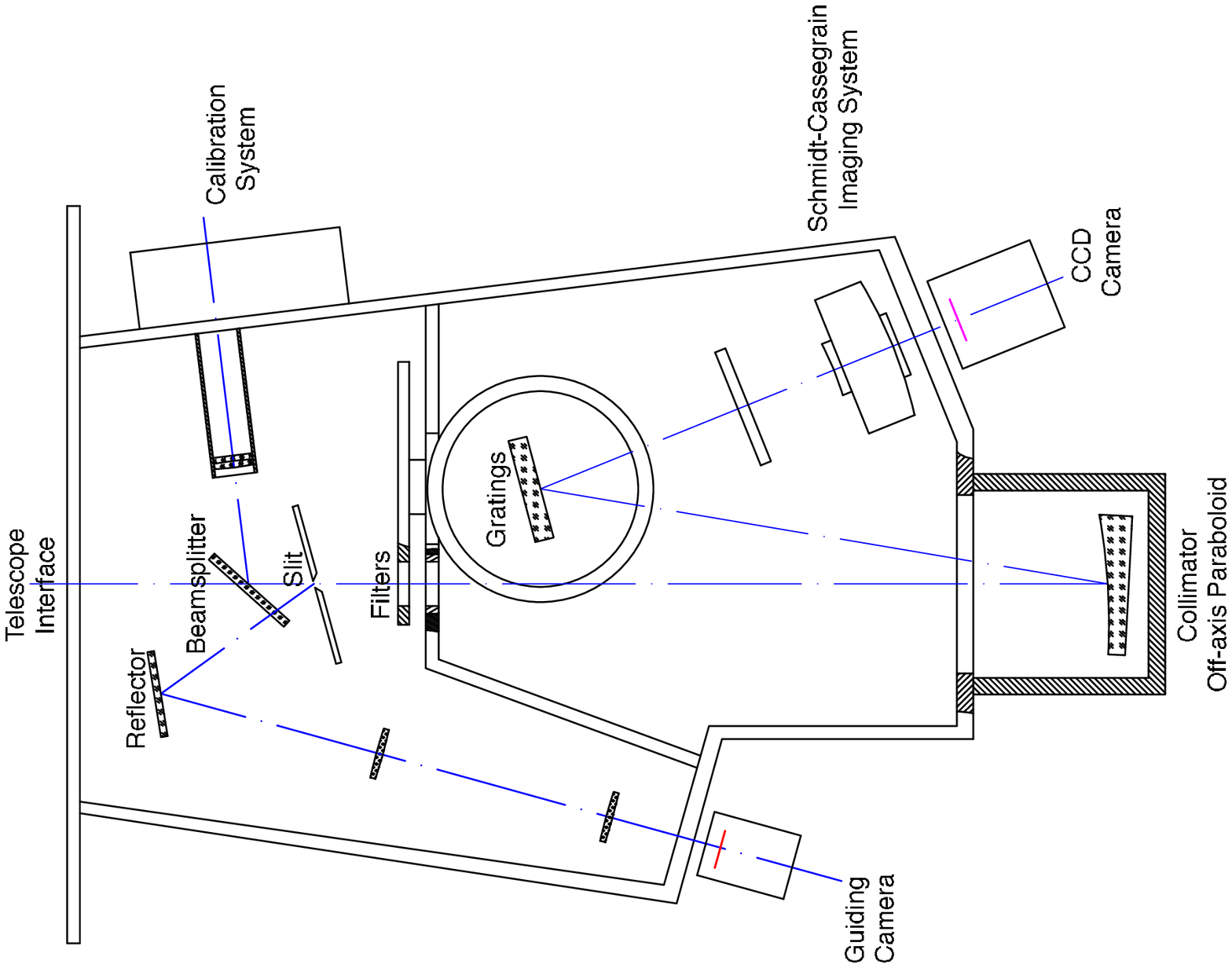}}
\caption[]{The optical layout of the OMR spectrograph.}
\label{fig11}
\end{figure}

The optical layout of the OMR is shown in Figure~\ref{fig11}. The system is 
composed of slit \& decker assembly, filters, spectrograph, calibration 
system, collimator, gratings, CCD camera, guiding CCD camera, console and data 
collecting system. 

The slit \& decker assembly is composed of decker, the slit width adjuster, 
calibration reflector, and driving device. The top surface of the slit is 
aluminized stainless steel with high reflecting power while the slit 
edge was polished to be very sensitive and should not be touched. The 
slit is tilted by $20^{\circ}$ with respect to telescope optical axis and the 
projected slit width 
can be adjusted remotely through the main console in the range of 0.05-1.0mm 
(corresponding to $0\farcs5-10\farcs6$), which is displayed on the main 
console via the voltage value. The size of the slit jaws is 32.8mm$\times$38mm 
and the effective length of the slit is 28.8mm (corresponding to $5'.1$). 
The slit jaw reflects the light to the guiding system. 

There are six positions on the filter wheel for the filters of Clear, Corning 
$4-71$, Schott $BNG-37$, $BG-39$, $GG-475$ and $RG-695$. All the filters have 
the same size of 25mm$\times$25mm. The central wavelengths of filters could be 
matched with the 1st or 2nd order of the spectrum by the softwares 
automatically.

There are three wavelength calibration lamps (Fe-Ne, Fe-Ar and He-Ar)
and a flat-fielding lamp for the calibration device. All of lamps can be 
controlled remotely. For the Fe-Ne lamp, a standard 1.5-inch ISTC Model
\#WL-22810 is supplied, and actually other types of lamps can 
be used as well, such as Hamamatsu and Starna. The maximum working current is 
20 mA and the normal working current is 10-12 mA. The Fe-Ar lamp ISTC Model 
is \#WL-22611. For the safety of He-Ar lamp, the DC power should 
be used for the supply. As for the flat-fielding lamp, the power is the 
standard 12V DC, 1.5A halogen tungsten lamp (Sylvania\#808-301550) and it
usually can be used for 1000-2000 hours. Meanwhile, there are four condensers 
for the calibration lamps, which are made of quartz or pyrex glass. The 
diameters are 25 mm and the focal lengths are $f=38$ mm. The field lens is 
made of quartz, with a diameter of 38 mm and a focal length of 64 mm.

The mirror of the collimator is an off-axis parabolic aluminized reflector, 
made of pyrex glass. The aperture of the mirror is $D=110$ mm and the focal 
length is $f=674$ mm. The off-axis angle is 8.1$^{\circ}$ and the focus can 
be adjusted remotely. 

At present there are six reflecting blazed gratings mounted on the OMR 
spectrograph. The gratings are made of aluminized pyrex glass and 
can be switched manually. The parameters of the blazed gratings are given in 
Table~\ref{t4.tab}, including the number of the gratings, the grooves, groove 
areas, reciprocal linear dispersions, dispersions, blaze wavelengths of the 
1st order spectra and the blaze angles. The overall working 
wavelength coverage of the gratings is $\sim$3700-10000 {\AA}. For the 
specific grating, the wavelength coverage is adjustable and it can be 
estimated through parameters of Table~\ref{t4.tab} and the size of the CCD. For 
instance, the wavelength coverage of gratings of 1200 lines/mm, is $\sim1380$ 
{\AA} while for the grating of 300 line/mm, it is $\sim5420$ {\AA}. The SPEC 
software, {which is the camera controlling software designed and installed with the 
PI Spec-10 CCD camera}, can recognize the gratings and the wavelength coverage 
can be adjusted via SPEC according to the requirements of the observers.

\begin{figure}
  \centerline{
    \includegraphics[scale=0.7,angle=0]{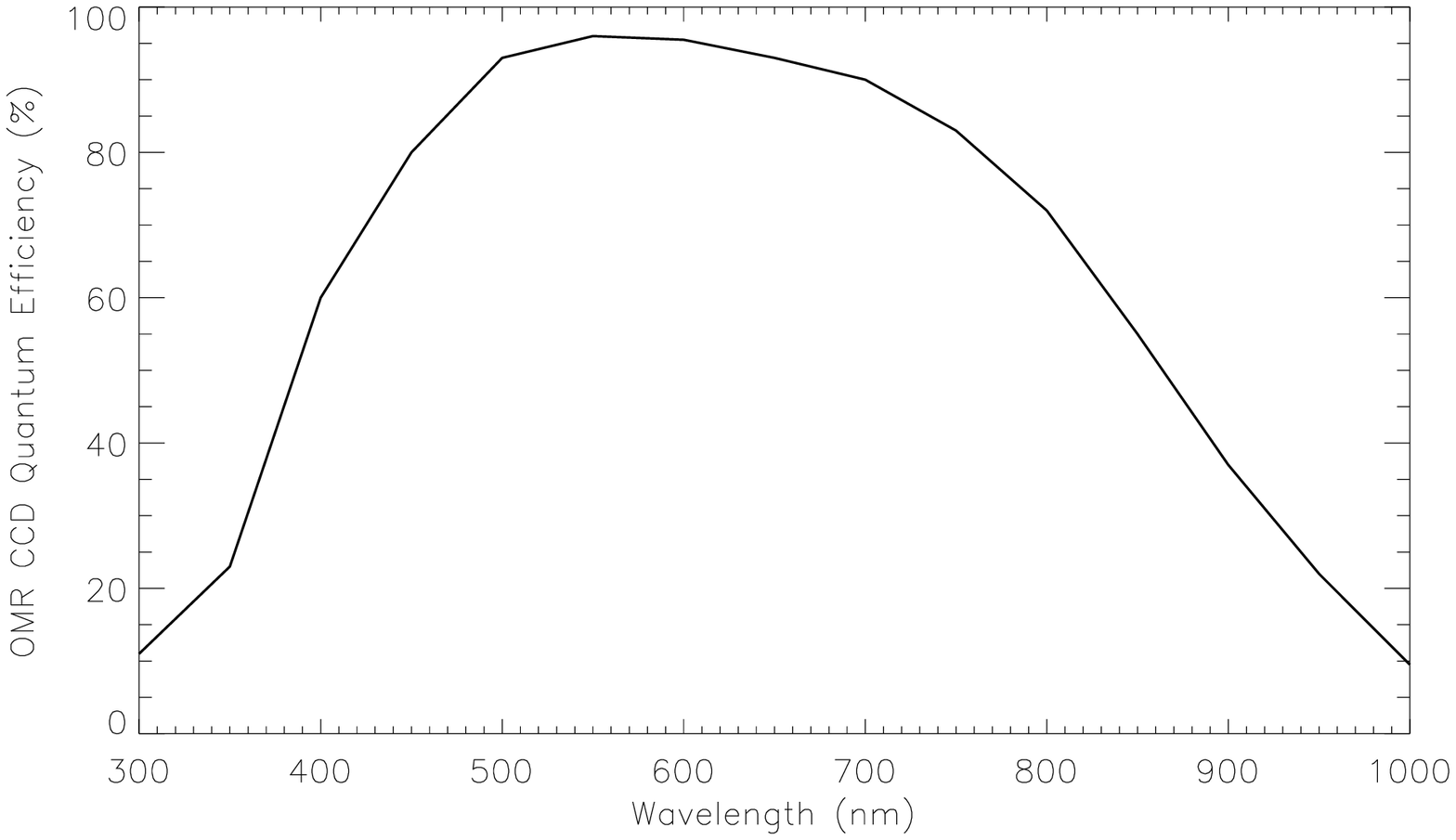}}
  \caption[]{The quantum efficiency of the CCD camera of the OMR Spec-10, which is midband.}
  \label{fig12}
\end{figure}

The camera is a Schmidt-Cassegrain system. The useful aperture is $D=100$ mm 
and the focal length is $f=150$ mm. The CCD mounted is a PI Spec-10 PIXIS 
$1340{\times}400$ scientific CCD detector, which delivers the highest 
sensitivity possible and $>16-$bit dynamic range for spectroscopy 
applications. The pixel size is 20$\times$20 $\mu$m pixels and the CCD size 
is $26.8{\times}8.0$ mm$^2$. The deepest cooling temperature is $-75^{\circ}$
C. The quantum efficiency of the CCD is shown in Figure~\ref{fig12}, which is
midband. 

The guiding system of the OMR is composed of a reflecting system and a CCD 
Camera. In the reflecting system, the slit reflecting mirror is aluminized and 
the r.m.s. of the surface flatness is 1/4 $\lambda$. For the transfer lens, 
the diameter is 50 mm, with focal length of $f=260$ mm. While for focal lens, 
the diameter is 36 mm, with focal length of $f=85$ mm. Both of the lenses have 
been coated. The CCD camera is an Alta U42 CCD, made by Apogee Imaging Systems,
Inc. The sensor is E2V CCD$42-40$ and the size is 2048$\times$2048 pixels. 
The Gain is 1.3e$^-$/count and the maximum digitized well capacity is 82k
e$^-$. The dark current is 0.39e$^-$/pixel/s.

\begin{table}[ht!!!]
\small
\centering
\begin{minipage}[]{100mm}
  \caption[]{The Current Parameters of the OMR Gratings.}\label{t4.tab}\end{minipage}
\tabcolsep 3mm
\begin{tabular}{ccccccc}
  \hline\noalign{\smallskip}
  No. & Grooves & Gro. Area & Rec. Lin. Disp. & Disp. & Blz. Wav. & Blz. Ang. \\
  & (l/mm)&(mm) & (\AA/mm) & (\AA/pix) &(1st order) (\AA)  & ($^{\circ}$)\\
  \hline\noalign{\smallskip}
  1 &  150  & $90\times90$   & 400 & 8.0 & 5000 &  2.2 \\
  2 &  300  & $90\times90$   & 200 & 4.0 & 8000 &  6.5 \\
  3 &  600  & $102\times102$ & 100 & 2.0 & 6500 & 11.3 \\
  4 &  1200 & $102\times102$ &  50 & 1.0 & 8000 & 28.7 \\
  5 &  300  & $102\times102$ & 200 & 4.0 & 4224 &  3.6 \\
  6 &  1200 & $102\times102$ &  50 & 1.0 & 4000 & 13.9 \\
  \noalign{\smallskip}\hline
\end{tabular}
\end{table}

\begin{figure}
\centerline{
  \includegraphics[scale=0.55,angle=0]{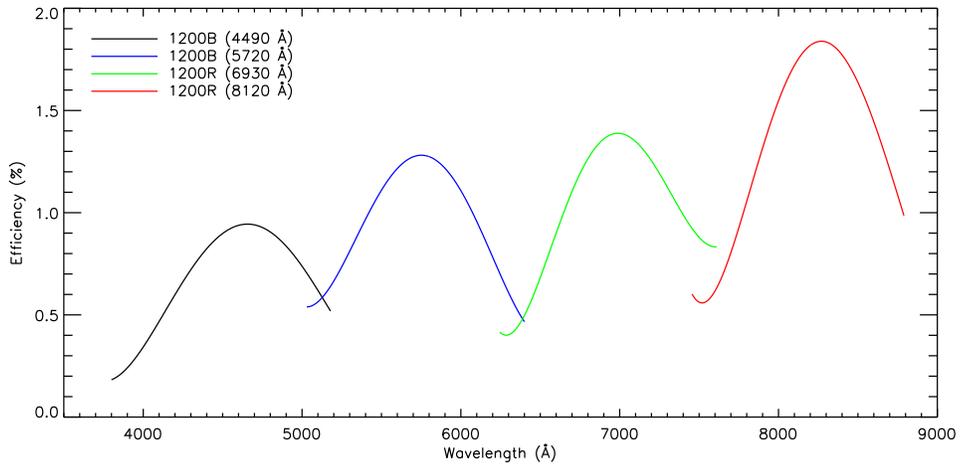}}
\caption[]{The total efficiency estimates with blazing gratings 1200B/1200R of 
  the OMR, in two different central wavelengths and wavelength coverages.}
\label{fig13}
\end{figure}

Similarly, the Landolt standard star Feige 66 was observed with two blazing 
gratings 1200B and 1200R of the OMR instrument and the total efficiencies 
(including the atmosphere extinction, the reflecting rate of mirrors, Q.E. 
of CCD, etc.) have been computed through Equation~\ref{eq1}. As described 
above, the central wavelength and coverage is adjustable, Figure~\ref{fig13} 
shows the total efficiencies of two gratings 1200B and 1200R, in two different
central wavelength and wavelength coverages. It can be seen that the 
efficiency ranges from $\sim 1\%$ to $\sim 2\%$, which is relatively lower than 
that of the BFOSC gratings.

Typically, for a star of $V=17.3$ mag in an exposure time of 1800s, the
signal-to-noise ratio is $SNR=12$ at wavelength of $\lambda$=5500 {\AA}; 
for a star of SDSS $g=19.8$ mag in an exposure time of 3600s, 
the signal-to-noise ratio is $SNR=5$ on average for the whole band.

\subsection{The fiber-fed High Resolution Spectrograph (HRS)}

Previously an echelle spectrograph was mounted on the Coud\'{e} focus of the 
telescope for high resolution spectroscopic observations. 
The special design of the optical system is to allow the Cassegrain
system and the Coud\'{e} system to share the same secondary mirror and there is
a relay mirror in the Coud\'{e} system. When switching between the two 
systems, the Coud\'{e} system can eliminate spherical aberration and coma 
aberration sufficiently just by slightly moving the secondary mirror 
\citep{szy89}. The resolution power was between R=16,000 to 170,000 (79 gr/mm)
in the blue beam from 330nm to 580nm
and R=13,000 to 170,000 (31.6 gr/mm) in the red beam from 520nm to 1100nm.
For an exposure time of 1 hour, the limiting magnitude could reach $V=9.5$ in
the red band and $V=7.2$ mag in the blue band with a signal-to-noise ratio
$S/N=100$ \citep{zhao00,zl01}.

\begin{figure}
\centerline{
  \includegraphics[scale=1.,angle=0]{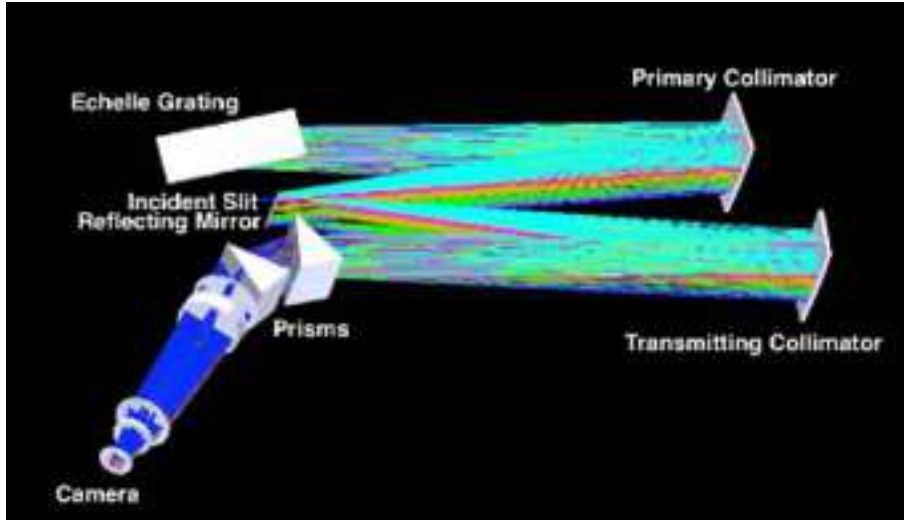}}
\caption[]{The optical layout and the light path of the fiber-fed HRS 
  spectrograph.}
\label{fig14}
\end{figure}

Since 2010, a new fiber-fed High Resolution Spectrograph (HRS) has
been developed by the Nanjing Institute of Astronomical Optics \&
Technology (NIAOT) to satisfy the scientific requirement of, for instance, 
exoplanets surveys, the study of stellar abundances, and stellar
magnetic activities. The optical layout and light path are shown in
Figure \ref{fig14}. The HRS system is available for the Cassegrain focus and
composed of four parts: 1) the Cassegrain connector of the telescope; 2) 
the fiber connector and the micro-optical module for focal-ratio changing; 
3) the main body of the spectrograph; 4) the data collecting part of the CCD 
camera.

The fiber-fed HRS observing mode could be switched to other instruments (such 
as BFOSC or OMR) conveniently through the Cassegrain focal interface
of the telescope within only a few minutes. The calibration system
of HRS, the $\rm I_2$ cell and its heating system, as well as the tip/tilt 
system are mounted on the interface of the spectrograph. The whole main
body is sealed in a protective box and the temperature variation is less 
than $r.m.s.=0.5^\circ$C for a week. At present, a $2\farcs4$ aperture fiber 
is configured and the tip/tilt system is working to improve the efficiency of
the system when the seeing is ideal. The fiber of $1\farcs6$ is at the
commissioning stage and it could be available very soon. An environmental 
controlling system can accurately keep the stability of both temperature and 
humidity for the HRS system. Furthermore, a sub-controlling system of 
pressure will be installed at this system in the future.

\begin{figure}
\centerline{
  \includegraphics[scale=0.7,angle=0]{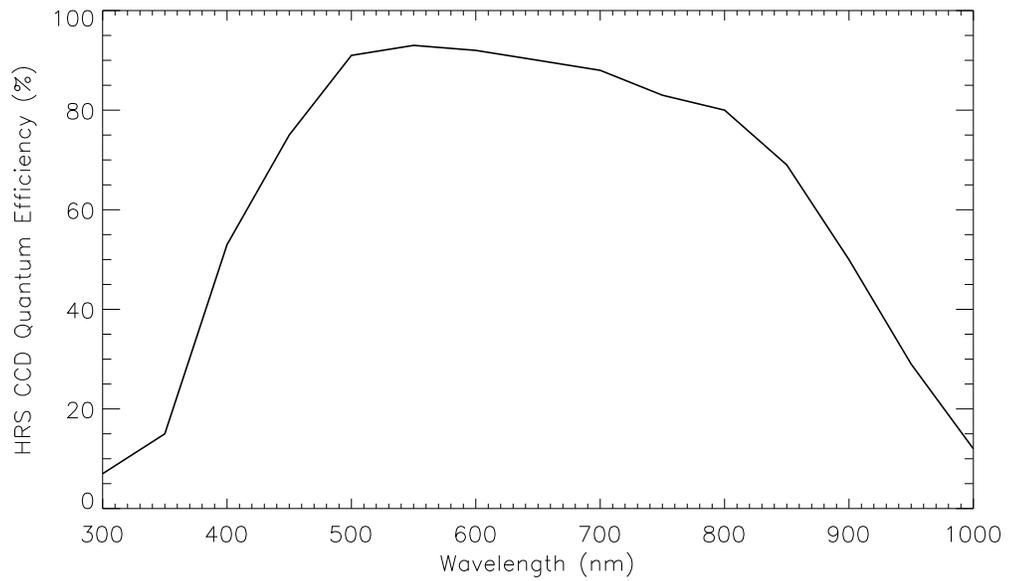}}
\caption[]{The quantum efficiency of the basic midband CCD camera used on the
  HRS at -100$^{\circ}$C, which is shown by the solid line}
\label{fig15}
\end{figure}

The basic parameters of HRS are listed in Table~\ref{t5.tab}. The working 
wavelength coverage is 360-1000 nm and the instrumental efficiency of 
the spectrograph is $\ge 34\%$ for the peak at  a wavelength coverage of 
640-790 nm ($RI$ band), and $\ge10\%$ for $\lambda > 4500$ {\AA} (the whole 
working band), based on the tests of November 23, 2010. The spectral resolution 
is 32,000-106,000 for the spectrograph and it is $R=49800$ at a fixed slit 
width of 0.19 mm (corresponding to 1\farcs8) based on the test of April 12, 
2011. The stability of 
the instrument for the velocity measurement is $r.m.s.=\pm6$ m/s and the 
temperature is quite stable even for two weeks. The CCD camera is a 
back-illuminated first-order red-sensitive E2V CCD 203-82. The size is 
$4096\times4096$ pixels with a pixel size of 12 $\mu$m. The typical quantum
efficiency of the CCD is $>90\%$ under the temperature of $-100^{\circ}$C in
the wavelength coverage of $\sim500-650$ nm, which is shown as the solid
line in Figure~\ref{fig15}.

The liquid nitrogen (LN) holding time of the system is $\sim20$ hours, and the 
cooling temperature is $-106^{\circ}$C. For $Gain=1.01$ e$^-$/ADU, the readout 
noise (RON) is 2.84 e$^-$ at the readout speed of 50k, 4.29 e$^-$ at readout 
speed of 100k and 7.88 e$^-$ at readout speed of 200k. The distance between 
the nearby orders of the spectrum is $\ge20$ pixels and the two fibers (
$2\farcs4$ aperture one and $1\farcs6$ aperture one) can work at the 
same time. The aperture of the fiber is 100 $\mu$m ($2\farcs4$) and the fiber
length is $\sim19$ m. The FOV for the guiding camera is $3'\times3'$ and in
the guiding plate the aperture of pupil is $4\farcs0$ in front of the fiber.
The temperature variation of the system is $r.m.s.=\pm0.05^{\circ}$C for a whole
night and $r.m.s.=\pm0.34^{\circ}$C for a week. The tip/tilt system is working
normally on the Cassegrain focus. The system is currently in its 
commissioning phase, and its radial velocity precision dose not yet reach 
its design goal of a few $\sim$cm/s. It is expected that this goal will be 
achieved once commissioning is complete. 

The guiding CCD camera is a GC1380, and the software $AVTUniCamViewer$ is 
used for CCD controlling, exposure time configuration and data collection. The 
telescope controlling software can monitor the guiding uncertainty of the  
guiding system.  

An astro-frequency comb calibration system, which is developed by Peking 
University (PKU), also was installed and is being tested on the 2.16-m
telescope. The full spectral wavelength range of the astro-frequency comb is 
160nm and the central wavelength is $640\pm20$ nm. The observed spacing of the 
comb teeth is 29.01 GHz. Once working normally, it can greatly improve the 
measuring accuracy of the stellar radial velocity. The system has significant 
advantages over the $\rm I_2$ cell \citep{wi10}. It consists of a series of 
discrete, equally spaced spectral lines with equal intensity and it is 
repeatable in long time scale. Right now, the system has been set up and the 
first spectrum simultaneously from the comb and the flat-fielding lamp has been
obtained during the engineering run. Meanwhile, a 25GHz AstroComb Optical 
Frequency calibration system manufactured by the MenloSystems company, also 
has been installed on the HRS instrument system recently. 
The spectral coverage is $\sim$450-720 nm, with the central wavelength of 
$540\pm30$ nm.
The flatness of the spectrum is r.m.s. $<5$ dB and the observed spacing of the 
comb teeth is 25 GHz. The luminous power of the system for the full spectral 
wavelength range is $>10\mu$W. The two new calibration systems are at the 
commissioning stage. The two new systems are supposed to be applied in 
the end of this year.

\begin{table}[ht!!!]
\small
\centering
\begin{minipage}[]{100mm}
  \caption[]{The Current Instrumental Parameters of the HRS.}\label{t5.tab}\end{minipage}
\tabcolsep 3mm
\begin{tabular}{ccc}
  \hline\noalign{\smallskip}
  Tech. Index & Parameters & Notes \\
  \hline\noalign{\smallskip}
  Wave. Cov.  & 3650-10,000 {\AA}      &  ThAr lamp  \\
  \hline
  Aper. of Fiber                 & $2\farcs4$ / $1\farcs6$ &  \\
  \hline
  Spec. Res.   &   32,000-106,000  & tested on Apr. 12, 2011   \\
  \hline
  Stab. of inst. &  $\pm6~m/s$        &                        \\
  \hline
  Effi.           & $34\%$ at peak value  &   \\
  & $\ge10$ \% for $\lambda > 4500$ {\AA}  & tested on Nov. 23, 2010   \\
  \hline
  & E2V 4k$\times$4k 12$\mu$m &        \\
  & Scientific chip, class 1   &   \\
  & On-chip Binning   &      \\
  CCD Camera & 50-200 KHz         &  2$\times$1 binning   \\
  & LN holding time: 20 hours &   mode suggested  \\
  & (w/LN auto-filling)  &      \\
  & LN cooling &  \\
  \hline
  FOV of Guid. Cam.   & $3'.3\times3'.3$    &    \\
  \hline
  Temp. var. / day & $\pm0.05^{\circ}$ &  \\
  \hline
  Temp. var. / week & $\pm0.34^{\circ}$ &  \\
  \noalign{\smallskip}\hline
\end{tabular}
\end{table}

In order to estimate the total efficiency (including the atmospheric 
extinctions and reflectivity of the primary/secondary mirrors, among others), 
a number of standard stars were observed with HRS on June 30, 2015.
By using Equation~\ref{eq1}, the final result shows that it is 
$>$2\% at 5500{\AA}.

\begin{figure}
\centerline{
  \includegraphics[scale=1.,angle=0]{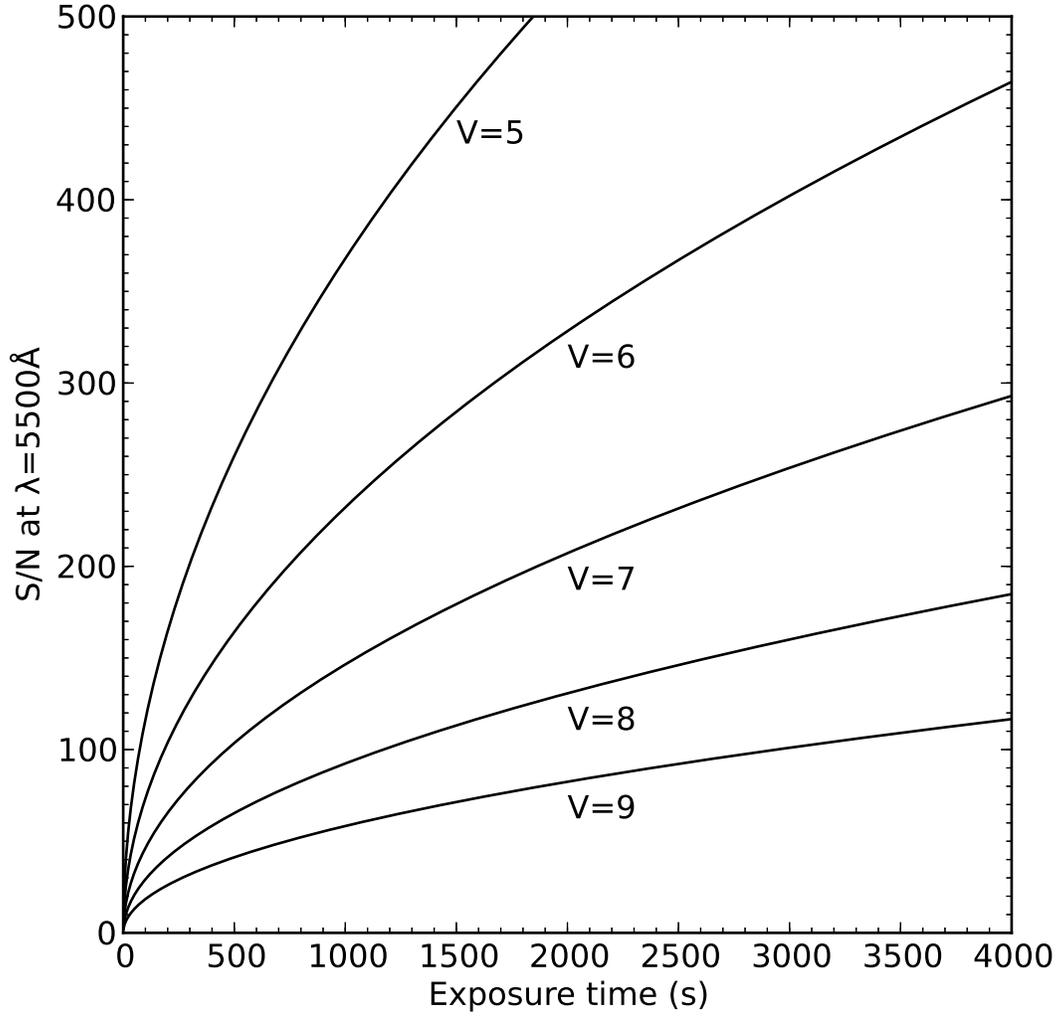}}
\caption[]{The signal-to-noise ratio at $\lambda$=5500 {\AA} as a function of 
  exposure time for stars from $V=5$ mag to $V=9$ mag by using the fiber-fed 
  HRS spectrograph.}
\label{fig16}
\end{figure}

In addition, the limiting magnitudes are estimated. Figure~\ref{fig16} 
is a plot of the signal-to-noise ratio (SNR) at $\lambda$=5500 {\AA} as a 
function of exposure time for the stars from $V=5$ mag to $V=9$ 
mag for the HRS system, based on the observing data. For a $V=9$ mag star, 
the typical signal-to-noise ratio $SNR=100$ for an exposure time of 
3000-3600 seconds. A few observing tests give the SNR in the typical observing 
conditions of 2012, for a star of $V=8.85$ mag with an exposure time of 
3600s, the signal-to-noise ratio is $S/N=80$ at wavelength of 
$\lambda=6000$ {\AA}, and for a star of $V=7.83$ mag in 2400s exposure, the 
signal-to-noise ratio is $SNR=150$ at wavelength of $\lambda$=6000 {\AA} 
(provided by Xiaoling Yang and Yuqin Chen).

\section{Scientific Projects}
\label{sci.sec}

Since 1989, when the 2.16-m telescope saw its first light, a great many 
scientific research projects have been carried out 
in various fields, including: the study of nearby galaxies (star
formation rates, gas and dust content), AGNs and their supermassive black
holes (SMBHs), quasars, stellar parameter determinations, exoplanets,
supernovae (SNe), gamma-ray bursts (GRBs), and follow-ups of tidal disruption
events. When LAMOST was built, the 2.16-m telescope has also 
been used for the LAMOST follow-up and stellar library observations.

Some of the scientific highlights obtained with the 2.16-m telescope
are as follows. In 1993, \citet{wh94} observed the spectra of supernova 1993J
with the  2.16-m telescope and found the blue-shifted oxygen lines only four
months after the optical discovery of the supernova, which is different from
the case of 1987A. $\sim100$ G-type giant stars of 
$V\sim6$ mag have been monitored with the Coud\'{e} HRS system \citep{zl01}. 
In a joint planet-search program between China and Japan by using the Xinglong 
 2.16-m telescope and Okayama Astrophysical Observatory (OAO), a number of 
substellar companions of intermediate-mass giant stars has been discovered 
by \citet{liu08,liu09}. Later, a brown dwarf companion candidate was discovered
by \citet{wang12} and a long-period giant planet has been discovered by 
\citet{wang14}, which is actually the first exoplanet discovered jointly 
with the Subaru telescope, the OAO 1.88m telescope and the Xinglong  2.16-m
telescope.

In recent years, in order to improve the efficiency of the telescope and to
maximize the scientific output, 7-8 key projects with the telescope
have been supported for 3-5 years. Each project usually owns $\sim20-40$ 
nights per year, which allows to carry out long-term observations and
large-sample surveys.
For instance, a number of quasars at intermediate redshift $2.2 < z < 3$ have 
been identified by \citet{wu11,wu13}. An H$\alpha$ imaging survey of 
$\sim$1400 nearby ALFALFA galaxies is carried out to study the star formation 
rate and the stellar distributions. From 1997 to 2002, the spectra of 
$\sim100$ blue compact galaxies (BCGs), which was the largest sample before 
SDSS, have been observed with the  2.16-m telescope, and the metallicity, 
extinction of dust, star formation rate of the sample galaxies have been 
derived by \citet{kc02,kong02,kong04,shi05,kong02}. A group of Peking 
University has discovered a high redshift quasar ($z=5.06$), J2202+1509, in 
November 2014 with the  2.16-m telescope, which is SDSS $i=18.79$ mag 
(the paper is in preparation). The redshift 
is derived from the emission lines, such as Ly$\alpha$ at wavelength of 1216 
{\AA}, and Nv at wavelength of 1240 {\AA}. This demonstrates that the 2.16-m 
telescope is able to be used to discover and study high redshift quasars.

\section{Observing Time Application}
\label{pro.sec}
Each year, the call for proposals begins around 20 October for the 
period of one month. Astronomers who are willing to use the telescope can 
submit proposals through the website http://astrocloud.china-vo.org before 
the deadline. After that, the proposals will be collected and 
reviewed by the Time Allocation Committee (TAC) of the 2.16m telescope. 
The probability of observing nights obtained for observers highly depends on 
the mark ranking. The ToO (Target-of-Opportunity) follow-up observations, 
like transient, SNe and GRBs, is supported, which are allowed to be applied 
and observed when the transient sources happen. The turn-around time 
(i.e. the time interval between a ToO alert and the start of the first 
observation) of the 2.16-m telescope depends on the system and configure 
being used when it happens. It concludes the following parts: the read-out 
time of last frame ($\sim$30-50 seconds); the time for changing the grisms and 
adjusting slit 
width and sometimes it even takes longer time when the grism is not on the 
spectragraph ($\sim$5-10 minutes); the pointing time of the telescope 
($\sim$1-5 minutes); the possible time for changing instrument and focusing 
($\sim$5-10 minutes), etc. Therefore, it takes $\sim$10-30 minutes 
in all for starting the ToO observations.

Since the first light of the 2.16-m telescope, there are astronomers from 
France, Japan, Taiwan, Hongkong, etc., beyond China Mainland, applied and 
used the telescope and publish papers. Until now, there are more than 150 
SCI papers published with the data obtained with the 2.16-m telescope since 
the first light. All these papers have been peer-reviewed and most of them 
were published on the high-impact magazines in astrophysics, e.g., $ApJ$, $AJ$,
$MNRAS$ and $A\&A$, and the impact factors are around $\sim4-6$, including the 
articles published in $Nature$. When the key projects are concluded, there 
will be numerous high impact journals published in the future.

\section{Summary and Discussion}
\label{sum.sec}

As a 2-meter class optical astronomical telescope in China, the 2.16-m telescope
of Xinglong Observatory of NAOC, plays an important role in observational 
astronomy today. There are currently three primary instruments, the BFOSC, OMR
and the fiber-fed HRS, available for the telescope users. 
When the 2.16-m telescope saw the first light in 1989, various scientific 
projects have carried out, based on observations of the telescope in the
research areas of nearby galaxies, AGN, supernovae, GRBs, exoplanets, Galactic 
stars, time-domain astronomy, and many other sciences. Since the LAMOST spectroscopic
survey started, the  2.16-m telescope has also been used for the LAMOST
follow-ups, in order to identify interesting objects, and for stellar library
observations. A great many remarkable studies have been done with the telescope,
including the spectroscopic research of the supernova 1993J \citep{wh94}.
In recent years, in order to improve the utilizing efficiency 
of the telescope and the scientific output further, 7-8 key observing 
projects have been set up, in which the observers own 20-40 nights/year 
to carry our long-term  observations and serial research. 

In the future, a number of potential new instruments will be available
to the  2.16-m telescope users to improve its observing abilities further, such
as an imaging photopolarimeter $RoboPol$ \citep[please see,][]{king14}, a
prototype of which is currently in use at the 1.3-m telescope of the Skinakas
Observatory in Crete, Greece, where it works well. A similar photo-polarimeter
will probably be installed at the 2.16-m telescope in the next one or two
years, which will be the first astronomical photo-polarimeter in China. It
could be used for polarization measurements, for instance of Seyfert galaxies,
blazars, and GRBs. Further, an intermediate resolution spectrograph (IRS) is
being investigated. If it is installed and committed in the future,
the projects of intermediate-resolution spectroscopic 
observations (e.g., X-ray binaries, bright stars, nearby galaxies, and 
a large number of fainter Galactic stars) could be carried out. In addition, 
the testing of the astro-frequency comb calibration system is at the 
commissioning stage.
When it is finished, the precision of the radial velocity measurements 
of Galactic stars could be improved greatly, even to a few cm s$^{-1}$.
Recently a number of adaptive optics experiments were performed on the 2.16-m 
telescope by engineers from various Chinese institutes, which can improve 
the spatial resolution and signal-to-noise ratio of targets to a large extent.
The exoplanet detection technology group from the Nanjing Institute of 
Astronomical Optics \& Technology, NAOC and the California State University, 
Northridge collaborate and built the High Performance Portable Adaptive Optics 
(HPAO) which is mounted on the Coud\'{e} focus of 2.16-m telescope and succeed 
after the testing observations. Although the limiting magnitude of the current 
HPAO system is faint to 3.8 mag in $H$ band, the system will be upgraded 
in the next step and the results can be improved significantly.

\begin{acknowledgements}
  This research was supported by the National Natural Science Foundation of
  China (NFSC) through grants 11003021, 11373003, 11273027, 11303042 and
  National Key Basic Research Program of China (973 Program) 2015CB857002.
  Z.F. acknowledges a Young Researcher Grant of the National Astronomical
  Observatories, Chinese Academy of Sciences. We thank Yuqin Chen, Liang Wang
  and Xiaoying Yang for providing the limiting magnitude measurements of the
  HRS instrument and thank Yingwei Chen for providing the picture of the
  2.16-m telescope.
\end{acknowledgements}

\appendix                  

\label{lastpage}
\end{document}